\documentclass[aps,pre,floats,amsmath,amssymb,showpacs,floatfix]{revtex4}

\usepackage{graphicx}
\usepackage{dcolumn}
\usepackage{psfrag}
\usepackage{float,color}
\newcommand{\ve}[1]{\ensuremath{\mbox{\boldmath$#1$}}}

\newcommand{\ku}{{{\rm Ku}}}
\newcommand{\st}{{{\rm St}}}

\newcommand{\Ordo}{\ensuremath{O}}

\newcommand\nn{\nonumber}
\newcommand{\ii}{{\ensuremath{\rm i}}}

\begin{document}
\title{Distribution of velocity gradients and rate of caustic formation
in turbulent aerosols at finite Kubo numbers}
\author{K. Gustavsson and B. Mehlig}
\affiliation{Department of Physics, Gothenburg University, 41296 Gothenburg, Sweden}

\begin{abstract}
In a one-dimensional model for a turbulent aerosol (inertial particles suspended
in a random flow) we compute the distributions of particle-velocity gradients
and the rate of caustic formation at finite but small Kubo numbers $\ku$,
for arbitrary Stokes numbers $\st$.
Our results are consistent with those obtained earlier
in the limit $\ku \rightarrow 0$ and $\st\rightarrow\infty$
such that $\ku^2 \st$ remains constant. We show how finite-time
correlations and non-ergodic effects influence the inertial-particle
dynamics at finite but small Kubo numbers.
\end{abstract}
\maketitle

\section{Introduction}
Turbulent aerosols (particles suspended in turbulent flows)
are ubiquitous in nature. Examples are microscopic rain droplets
suspended in the turbulent air flow of cumulus clouds \cite{Sha03},
and the motion of dust particles suspended in the gas
surrounding a growing star (see \cite{Wil08} and references
cited therein).  The motion of small, non-interacting  particles suspended in a fluid is commonly modeled by the equation of motion:
\begin{equation}    
\label{eq:1}
\dot {\ve r} = \ve v\,,\quad \dot{\ve v} = \gamma (\ve u-\ve v)\,.
\end{equation}
Here $\ve r$ is the particle position, $\ve v$ its velocity,
$\ve u(\ve r,t)$ is the velocity field of the flow, and
$\gamma$ is the Stokes damping rate. Eq. (\ref{eq:1}) assumes
that the particle Reynolds number is small, that Brownian
diffusion of the particles is negligible, and that the inertia
of the displaced fluid can be neglected.

An important dimensionless parameter
of the problem is the ratio of the Stokes damping time $\gamma^{-1}$ to the
correlation time $\tau$ of the underlying flow at small length scales.
This ratio is commonly referred to as the Stokes number, $\st = (\gamma \tau)^{-1}$. When the Stokes number is small, the particles are advected by the
flow $\ve u(\ve r,t)$. When the Stokes number is large, by contrast,
particle inertia becomes important, allowing the particles to detach from
the flow. It has been observed in direct numerical simulations
of particles suspended in turbulent flows that particle inertia may give
rise to large relative velocities between aerosol particles
\cite{Wang,Collins,Bec}. Large
relative velocities on small length scales in turn imply
large collision rates. It is of great importance to quantitatively
understand and to parameterise the $\st$-dependence of collision rates of
particles suspended in turbulent flows,
because collision rates determine the stability of turbulent aerosols.

While the dynamics of particles advected in turbulent flows
is very well understood \cite{Fal01}, inertial
particle dynamics is more difficult to treat, and requires
approximations. One possibility is to expand around
the advective limit (assuming small Stokes numbers)
\cite{Bal01,Wil07}. However, this method does not capture the occurrence of singularities
in the particle dynamics that give rise
to large relative velocities at small separations \cite{Fal02,Wil03,Wil06}.
These singularities occur when the phase-space manifold
(describing the dependence of
particle velocity upon particle position)
folds over.
In \cite{Wil05} these singularities were
identified as \lq caustics' analogous to light patterns on the bottom of a swimming
pool on a sunny day.  In the fold region between caustics, the velocity field at a given point
in space becomes multi-valued, giving rise to large velocity
differences between nearby particles.

\citet{Wil03} considered a limit of
the problem where the flow fluctuates rapidly. The relevant
dimensionless parameter is the \lq Kubo number' $\ku = u_0 \tau/\eta$
where $u_0$ is the typical size of the flow velocity, and $\eta$ is
its correlation length. The Kubo number characterises the fluctuations
of $u(\ve r,t)$. In the limit $\ku\rightarrow 0$, the
suspended particles experience the flow
as a white-noise signal, and their dynamics is \lq ergodic':  
the fluctuations of $\ve u(\ve r(t),t)$ (and its derivatives) along a particle
trajectory $\ve r(t)$ are indistinguishable from the fluctuations
of $\ve u(\ve r_0,t)$ at a fixed point $\ve r_0$. In this case
the instantaneous configuration of the flow field is irrelevant to the dynamics
of the suspended particles.
In turbulent flows, the Kubo number is of order unity. Nevertheless,
this approach has in the past yielded important insights into the dynamics of turbulent aerosols
\cite{Wil06,And07}. In the limit of $\ku\rightarrow 0$, $\st\rightarrow \infty$
(such that $\epsilon^2 = 3 \ku^2 \st$ remains constant) it is possible to
compute the fluctuations of particle-velocity gradients (that characterise spatial
clustering of the suspended particles), the rate of caustic formation, and the distribution
of relative velocities.

In \cite{Gus11,Wil11} it was recently shown how to compute Lyapunov exponents characterising spatial clustering
of inertial particles at finite Kubo numbers. It was found that two mechanisms for spatial clustering
compete (\lq preferential concentration' \cite{Max87} and \lq multiplicative
amplification' \cite{Gus11}).

This raises the question of how
the distribution of particle-velocity gradients and the rate of caustic formation
are modified at finite Kubo numbers.
This question is addressed in the present paper. For a one-dimensional
model of inertial-particle dynamics we compute the distribution
of particle-velocity gradients and the rate of caustic formation at finite Kubo numbers.

The remainder of this paper is organised as follows. In Section \ref{sec:model},
the model is introduced. The distribution of particle-velocity gradients
is computed in Section \ref{sec:PzA}, by means of a perturbation expansion
in $\ku$. This expansion describes the body of the distribution well,
but does not capture its tails (related to the formation of caustics).
The rate of caustic formation is computed in Section \ref{sec:caustic} using
a WKB approximation valid for small Kubo numbers and arbitrary
Stokes numbers. As in the white-noise limit, the problem can
be mapped to an escape problem, but now in the presence of coloured
noise \cite{Kane}. Finally, conclusions are summarised in Section \ref{sec:conc}.

\section{Model}
\label{sec:model}
Understanding inertial particle dynamics at
finite Stokes and Kubo numbers is a difficult problem.
In this paper we therefore analyse inertial particle dynamics
in one spatial dimension. In this case Eq.~(\ref{eq:1}) takes the form:
\begin{equation}
\label{eq:1d}
\dot x = v\,,\quad \dot v = \gamma (u(x(t),t)-v) \,.
\end{equation}
To simplify the problem further, we take $u(x,t)$ to be
a random function with correlation time $\tau$, correlation
length $\eta$, and typical fluctuation size $u_0$. We write $u(x,t) = \nabla \phi(x,t)$ where $\phi$
is a homogeneous random function with zero mean and correlation function
\begin{equation}
\langle \phi(x,t) \phi(0,0) \rangle =u_0^2\eta^2e^{-x^2/(2\eta^2)-|t|/\tau}\,.
\end{equation}
This one-dimensional model was analysed by \citet{Wil03} who computed
the distribution of velocity gradients $z= \partial v/\partial x$
and the rate of caustic formation in the limit
$\ku\rightarrow 0$, $\st \rightarrow \infty$
so that $\epsilon^2 = 3 \ku^2 \st$ remains constant.
In this limit, $z$ satisfies the Langevin equation
\begin{equation}
\dot z = \gamma(A-z)-z^2
\end{equation}
where $A=\partial u/\partial x$ is Gaussian white noise with
zero mean and correlation function $\langle A(t) A(t') \rangle = 2\gamma\epsilon^2  \delta(t-t')$.
The steady-state solution of the corresponding Fokker-Planck equation
for the distribution of $z$ is \citep{Wil03}:
\begin{equation}
P(z) = -\frac{J}{\gamma^3\epsilon^2}{\rm e}^{-S(z)}\int_{-\infty}^z{\rm d}z'{\rm e}^{S(z')}\,,
\label{eqn:distribution_z_WN}
\end{equation}
where $S(z)=z^3/(3\gamma^3\epsilon^2)+z^2/(2\gamma^2\epsilon^2)$ and $J<0$ is a constant
probability current towards negative values of $z$.
Caustics occur as $z$ passes from $-\infty$ to $+\infty$ in a finite time. This happens at a rate determined by $|J|$.
The rate of caustic formation is given by:
\begin{equation}\label{eq:J}
\frac{|J|}{\gamma} = \frac{1}{2\pi}{\mathcal Im}\Big[\frac{ {\rm Ai}'(y) }{\sqrt{y} {\rm Ai}(y)}\Big]\Bigg|_{y=(-1/(8\,\epsilon^2))^{2/3}}
\sim \frac{1}{\sqrt{2\pi}} {\rm e}^{-1/(6\epsilon^2)}
\end{equation}
where the last expression is valid for small $\epsilon$.
Eq.~(\ref{eq:J}) is equivalent to the formula derived by \citet{Wil03}.
As caustics form, excursions to large values of $|z|$ result in slowly decaying tails of the distribution of $z$.
It is easily seen that the distribution (\ref{eqn:distribution_z_WN})
exhibits power-law tails on the form $P\sim C/|z|^2$ for large values of $z$.

\section{Distribution of $z$ and $A$ at finite Kubo numbers}
\label{sec:PzA}
\begin{figure}
\psfrag{y}{\hspace*{-1.5cm} $P(A)/P_0(A)-1$}
\psfrag{x}{\raisebox{-2mm}{ $A/\sqrt{3}$}}
\includegraphics[width=6cm]{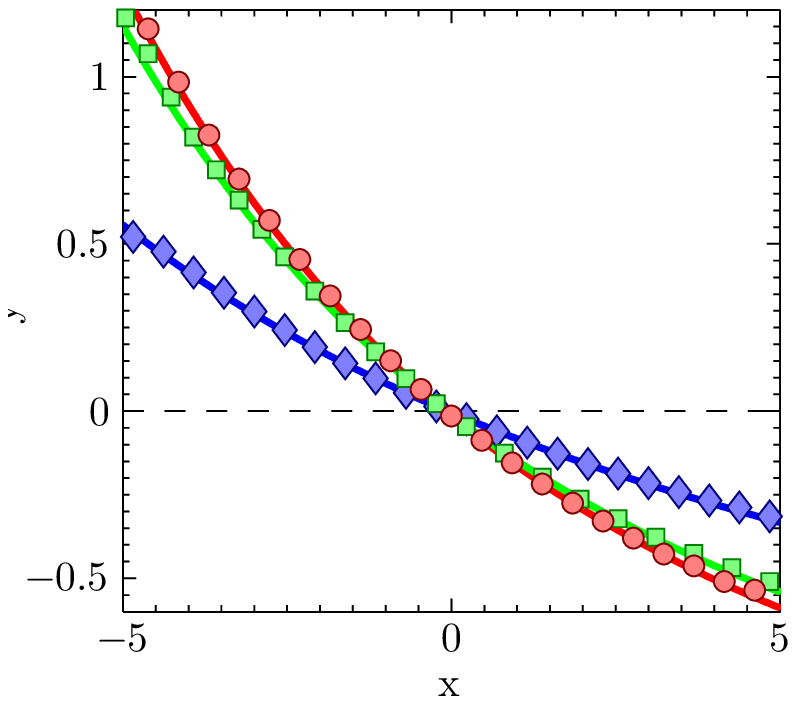}\hspace*{1cm}
\psfrag{y}{$P(z)$}
\psfrag{x}{\raisebox{-2mm}{\hspace*{-1.4cm} $z\sqrt{(1+\st)/3}$}}
\includegraphics[width=6cm]{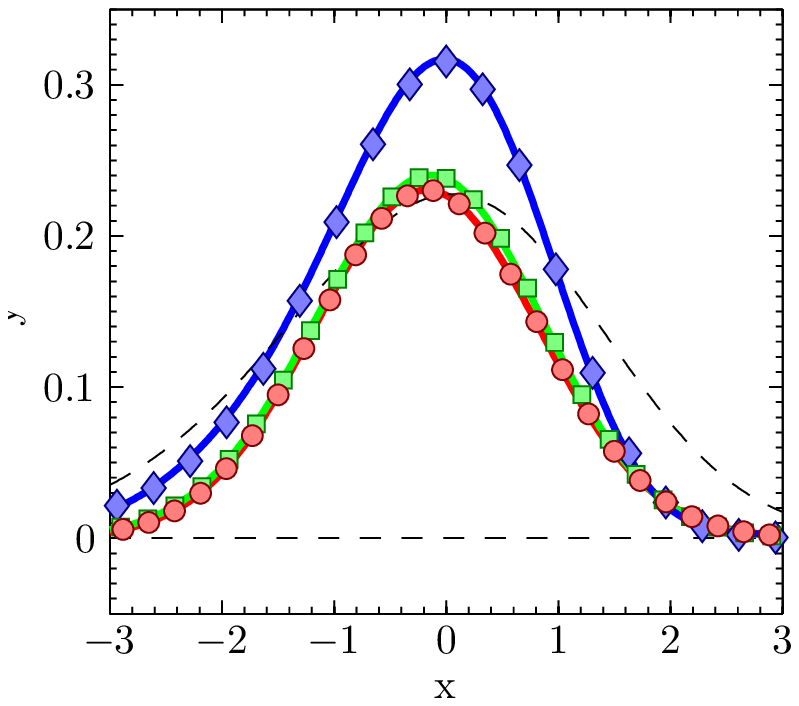}
\caption{\label{fig:dists} Distributions of $z$ and $A$.
Left: difference between $P(A)$ and $P_0(A)$. Shown are
results of numerical simulations of the model described
in Section \ref{sec:model} for $\ku=0.1$,
$\st=0$ (red $\circ$), $ \st=0.1$ (green $\Box$), and $\st = 1$ (blue $\Diamond$).
Also shown are results of the $\ku$-expansion, Eq.~(\ref{eqn:distA_1d_final}) extended
to fourth order in $\ku$.  Right:  distribution $P(z)$, same parameters as left.
Also shown are the $\ku$-expansion Eq.~(\ref{eqn:distz_1d_final}) extended
to second order in $\ku$.  Also shown is the white-noise result (\ref{eqn:distribution_z_WN}), dashed line. }
\end{figure}
In this section we show how to compute the steady-state distributions
of particle- and flow-velocity gradients ($z = \partial v/\partial x$ and
$A = \partial u/\partial x$ respectively) at finite Kubo numbers.
The method is based on a perturbative calculation of the moments
of $z$ and $A$. It works well when $|z|$ and $|A|$ are not too large,
but fails in the tails of the distribution. The tails of the distribution
of $z$ are due to the formation of caustics, not described by the
perturbation theory employed in this section. The formation of
caustics at finite Kubo numbers is discussed in Section~\ref{sec:caustic}.

We introduce dimensionless variables:
$t=t'\tau$, $x= x'\eta$, $v= v' u_0$, and $u= u' u_0$, where $\tau$, $\eta$ and $u_0$
are characteristic time-, space- and velocity scales discussed above. Dropping
the primes to simplify the notation, Eq.~(\ref{eq:1d}) becomes
\begin{align}
\dot{x}&=\ku\, {v}\,,\quad  \dot{v}=\frac{1}{\st}\, (u-v)\,.
\label{eq:1ddd}
\end{align}
Eq.~(\ref{eq:1ddd}) is difficult
to solve because $u$ depends non-linearly upon $x(t)$.
In the following we describe an approximate solution in terms of 
a perturbation expansion in powers of $\ku$. In \cite{Gus11} this method
was used to compute the Lyapunov exponents of inertial particles
suspended in one- and two-dimensional random flows at finite
Kubo numbers.

\subsection{Method}
The method is based on an expansion of the implicit solution of (\ref{eq:1ddd}):
\begin{equation}
x(t)=x_0+\ku\st v_0\left(1-e^{-t/\st}\right)+\frac{\ku}{\st}\int_{0}^{t} {\rm d}t_1\int_{0}^{t_1}{\rm d}t_2e^{-(t_1-t_2)/\st}u(x(t_2),t_2)\,.
\label{eqn:rsolution}
\end{equation}
Here $x_0\equiv x(0)$ is the initial particle position, and $v_0\equiv v(0)$ is the initial particle velocity. 
Now consider the difference $\delta x(t)=x(t)-x_0$ between the actual trajectory of a particle and its initial position.
Note that $\delta x(t)$ is proportional to $\ku$ and can therefore be considered small provided $\ku$ is small enough.
In this case, one may expand  $u(x(t),t)$ in powers of $\delta x(t)$:
\begin{equation}
u(x(t),t)=u(x_0,t)+\frac{\partial u}{\partial x}(x_0,t)\delta x(t)+\frac{1}{2}\frac{\partial^2 u}{\partial x^2}(x_0,t)\delta x(t)\delta x(t)+\dots\,.
\label{eqn:uexpansion}
\end{equation}
Inserting $\delta x(t) = x(t)-x_0$ from Eq.~(\ref{eqn:rsolution}) into  Eq.~(\ref{eqn:uexpansion}) yields to second order in $\ku$
\begin{align}
&u(x(t),t)
=u(t)
+\ku\st v_0\frac{\partial u}{\partial x}(t)\left(1-e^{-t/\st}\right)
+\frac{\ku}{\st}\frac{\partial u}{\partial x}(t)\int_{0}^{t} {\rm d}t_1\int_{0}^{t_1}{\rm d}t_2e^{-(t_1-t_2)/\st}u(x(t_2),t_2)\nn\\
&+\frac{1}{2}\ku^2\st^2 v_0^2\frac{\partial^2 u}{\partial x^2}(t)\left(1-e^{-t/\st}\right)^2
+\ku^2 v_0\frac{\partial^2 u}{\partial x^2}(t)\left(1-e^{-t/\st}\right)\int_{0}^{t} {\rm d}t_1\int_{0}^{t_1}{\rm d}t_2e^{-(t_1-t_2)/\st}u(x(t_2),t_2)\nn\\
&+\frac{1}{2}\frac{\ku^2}{\st^2}\frac{\partial^2 u}{\partial x^2}(t)\int_{0}^{t} {\rm d}t_1\int_{0}^{t} {\rm d}t_2\int_{0}^{t_1}\!\!\!\!\!{\rm d}t_3\int_{0}^{t_2}\!\!\!\!\!{\rm d}t_4e^{-(t_1+t_2-t_3-t_4)/\st}u(x(t_3),t_3)u(x(t_4),t_4)
+\Ordo(\ku^3)\,.
\label{eqn:uexpansion_Ku2_tmp}
\end{align}
Note that $u(x(t),t)$ occurs on both sides of this equation. By iteratively substituting $u(x(t),t)$ we find:
\begin{align}
&u(x(t),t)=
u(t)
+\ku\st v_0\frac{\partial u}{\partial x}(t)\left(1-e^{-t/\st}\right)
+\frac{\ku}{\st}\int_{0}^{t} {\rm d}t_1\int_{0}^{t_1}{\rm d}t_2e^{-(t_1-t_2)/\st}\frac{\partial u}{\partial x}(t)u(t_2)\nn\\
&+\ku^2v_0\int_{0}^{t} {\rm d}t_1\int_{0}^{t_1}{\rm d}t_2e^{-(t_1-t_2)/\st}\frac{\partial u}{\partial x}(t)\frac{\partial u}{\partial x}(t_2)\left(1-e^{-t_2/\st}\right)\nn\\
&+\frac{\ku^2}{\st^2}\int_{0}^{t} {\rm d}t_1\int_{0}^{t_1}{\rm d}t_2\int_{0}^{t_2} {\rm d}t_3\int_{0}^{t_3}{\rm d}t_4e^{-(t_1-t_2+t_3-t_4)/\st}\frac{\partial u}{\partial x}(t)\frac{\partial u}{\partial x}(t_2)u(t_4)\nn\\
&+\frac{1}{2}\ku^2\st^2 v_0^2\frac{\partial^2 u}{\partial x^2}(t)\left(1-e^{-t/\st}\right)^2
+\ku^2 v_0\left(1-e^{-t/\st}\right)\int_{0}^{t} {\rm d}t_1\int_{0}^{t_1}{\rm d}t_2e^{-(t_1-t_2)/\st}\frac{\partial^2 u}{\partial x^2}(t)u(t_2)\nn\\
&+\frac{1}{2}\frac{\ku^2}{\st^2}\int_{0}^{t} {\rm d}t_1\int_{0}^{t} {\rm d}t_2\int_{0}^{t_1}\!\!\!\!\!{\rm d}t_3\int_{0}^{t_2}\!\!\!\!\!{\rm d}t_4e^{-(t_1+t_2-t_3-t_4)/\st}\frac{\partial^2 u}{\partial x^2}(t)u(t_3)u(t_4)
+\Ordo(\ku^3)\,.
\label{eqn:uexpansion_Ku2}
\end{align}
This equation expresses the velocity field $u(x(t),t)$ at the particle position $x(t)$ in terms of $u(t) \equiv u(x_0,t)$ and its spatial derivatives. In the same manner higher-order contributions in $\ku$ can be included. Eq.~(\ref{eqn:uexpansion_Ku2}) constitutes an expansion of $u(x(t),t)$ in powers of $\ku$.  The products of the form $\cdots \partial^2u/\partial x^2(t) u(t_3) u(t_4) \cdots $
occurring in the integrands on the right-hand side of  Eq.~(\ref{eqn:uexpansion_Ku2}) can be averaged
using the known statistical properties of the velocity field $u$.

We note that in a similar manner, one may expand the strain $A\equiv\partial u/\partial x$ (as well as
higher order derivatives of $u$) along a trajectory.
The result is similar to the above,
but the leftmost factors $\frac{\partial^nu}{\partial x^n}(t)$, $n=0,1,\dots$ in (\ref{eqn:uexpansion_Ku2}) 
are replaced by $\frac{\partial^n A}{\partial x^n}(t)$.
Similarly particle-velocity gradients $z = \partial v/\partial x$ can be averaged along particle trajectories.

As outlined in the introduction, the subject of this paper are steady-state distributions of $z$ and $A$. Below they are computed via steady-state average
of moments of $z$ and $A$. We denote the steady-state average of a quantity $F$ by
\begin{equation}
\overline{F(x(t),t)}=\lim_{T\to\infty}\frac{1}{T}\int_0^T{\rm d}tF(x(t),t)=\lim_{T\to\infty}\frac{1}{T}\int_0^T{\rm d}t\langle F(x(t),t)\rangle\,.
\label{eqn:timeavg}
\end{equation}
In the limit of $T\rightarrow \infty$, a long trajectory
can be viewed as a concatenation of many long trajectories
which give rise to an ensemble (the average over which is
denoted by $\langle\cdots\rangle$)
over initial configurations $\{x(0),v(0),u(0),\partial^i_x u(0),\dots\}$.
In this limit, we expect all information about the initial configuration to be lost. At this point we therefore set $x(0)=v(0)=0$ in (\ref{eqn:uexpansion_Ku2}).  We also set
the initial velocity (and its spatial derivatives) to zero, $u(0)=0$. That the solution becomes independent of the initial configuration (and hence independent
of the distribution from which we draw the initial configuration) is explicitly shown for a number of examples in \cite{Gus12a}.

In summary, the expansion described above allows to compute moments of $u$, $v$, $z = \partial v/\partial x$, and $A = \partial u/\partial x$.
Up to this point no assumption is made about the statistical properties of the velocity field $u$. In the following we take $u$ to be Gaussian distributed;
this allows us to use Wick's theorem in evaluating the terms in the perturbation expansion. For the particular model described in section
\ref{sec:model} we have (for  $m,n=0,1,2,\dots$):
\begin{align}
\left\langle\frac{\partial^n u}{\partial x^m}(x(0),t_1)\frac{\partial^n u}{\partial x^n}(x(0),t_2)\right\rangle=
\left\{
\begin{array}{ll}
(-1)^{(n-m)/2}(m+n+1)!!e^{-|t_1-t_2|} & \mbox{if }m+n\mbox{ even}\,,\cr
0 & \mbox{otherwise}\,.
\end{array}
\right.
\label{eqn:correlation_uu_1d}
\end{align}
The method described above
was used to compute the Lyapunov exponents for inertial particles suspended in random flows
in one and two spatial dimensions in \cite{Gus11}. In the following we show how to compute
the distributions of $z$ and $A$.

\subsection{Moments and distributions of $z$ and $A$}
Using the method described in the previous subsection, we obtain
the following expressions for
the steady-state averages of the moments of $A$ (for $n=0,1,\dots$):
\begin{align}
\label{eqn:moments_A_odd}
\overline{A(x(t),t)^{2n+1}}&=-3^{n+1}(2n+1)!!\frac{\ku}{1+\st}\,,\\
\overline{A(x(t),t)^{2n}}&=3^n(2n-1)!!\bigg[1+3n\frac{\ku^2(1+3\st)}{(1+\st)^2(1+2\st)}\bigg]\,.
\label{eqn:moments_A_even}
\end{align}
The distribution $P(A)\equiv  P(A(x(t),t))$
is found by relating the moments of $A$ to the Fourier transform $\widetilde P(\alpha)$ of $P(A)$:
\begin{align}
\langle A^n\rangle&=\int_{-\infty}^\infty{\rm d}A A^n P(u,A)
=\left.\frac{1}{(-\ii)^{n}}\frac{\partial^n}{\partial\alpha^n}\int_{-\infty}^\infty{\rm d}u\int_{-\infty}^\infty{\rm d}A e^{-\ii\alpha A} P(A)\right|_{\alpha=0}
=\left.\frac{2\pi}{(-\ii)^{n}}\frac{\partial^n}{\partial\alpha^n}\widetilde P(\alpha)\right|_{\alpha=0}\,.
\end{align}
Inserting this expression into the Taylor expansion of $\widetilde P(\alpha)$ and using the moments in (\ref{eqn:moments_A_odd}) and (\ref{eqn:moments_A_even}) we find
\begin{align}
&\widetilde P(\alpha)=\left.\sum_{n=0}^\infty\frac{\alpha^n}{n!}\frac{\partial^n}{\partial\alpha^n}\widetilde P(\alpha)\right|_{\alpha=0}
=\frac{1}{2\pi}\sum_{n=0}^\infty\frac{(-\ii\alpha)^n}{n!}\langle A^n\rangle =\frac{1}{2\pi}\bigg[1 +3\ii\frac{\ku}{1+\st}\alpha-
9\alpha^2\frac{\ku^2(1+3\st)}{2(1+\st)^2(1+2\st)}\bigg] e^{-3\alpha^2/2}\,.
\end{align}
Finally, the inverse transform yields  the desired distribution:
\begin{equation}
\label{eqn:distA_1d_final}
P(A)=\frac{1}{\sqrt{6\pi}}\left[1-\frac{A\ku}{1+\st}+\frac{(A^2-3)\ku^2(1+3\st)}{2(1+\st)^2(1+2\st)}\right]e^{-A^2/6}\,.
\end{equation}
In the limit of $\ku\rightarrow 0$ (or $\st\rightarrow \infty$) 
this distribution converges to  $P_0(A)=\exp[-A^2/6]/\sqrt{6\pi}$, the distribution of flow-velocity
gradients evaluated at fixed position $x_0$. At finite values of $\ku$, by contrast,
the distribution of $A$ is shifted towards negative values of $A$
(to first order in $\ku$). To this order, the mean of $A$ is
\begin{equation}
\overline{A} = \frac{-3\ku}{1+\st}\,.
\end{equation}
The fact that $\overline{A}$ is not zero is a consequence of preferential concentration \cite{Gus11}.
This average was computed by \citet{Wil11} in the advective limit ($\st=0$) and used
to determine the maximal Lyapunov exponent for inertial particles
in one spatial dimension at finite Kubo numbers. A corresponding average 
figures in a one-dimensional model where the fluid-velocity gradients $A$
fluctuate according to a telegraph process \cite{Fal07}.

Moreover, to second order in $\ku$ the width of the distribution $P(A)$ increases.
The difference between $P(A)$ and $P_0(A)$ is shown and compared to results of numerical simulations
of the model described in Section \ref{sec:model} in Fig.~\ref{fig:dists}.

We now turn to the distribution of $z$. Motivated by the form of (\ref{eqn:distA_1d_final}) we try the following ansatz for $P(z)$:
a polynomial in $z$ times the $\ku=0$-distribution $P_0(z)$. For small values of $\ku$ and $|z|$, $P_0(z)$ is
approximately Gaussian with zero mean and variance $3/(1+\st)$.
From the following expressions for the steady-state moments of $z$
\begin{align}
\overline{z}&=-3\ku\,,\\
\overline{z^3}&=-9\ku\frac{6 + 33 \st + 39 \st^2 + 10 \st^3}{(1 + \st)^2 (2 + \st) (1 + 2 \st)}\,,\\
\overline{z^5}&=-135\ku\frac{6 + 45 \st + 57 \st^2 + 14 \st^3}{(1 + \st)^3 (2 + \st) (1 + 2 \st)}\,,\\
\overline{z^7}&=-8505\ku\frac{2 + 19\st + 25\st^2 + 6\st^3}{(1 + \st)^4 (2 +\st) (1 + 2\st)}\,,
\end{align}
we can determine the coefficients of the polynomial multiplying $P_0(z)$. This gives to first order in $\ku$:
\begin{align}
&P(z)=\sqrt{\frac{1+\st}{6 \pi }}e^{-z^2(1+\st)/6}\left[1-\ku\frac{2+\st-2\st^2}{(2+\st)(1+2\st)}z
-\frac{\ku}{9}\frac{\st(1+\st)(6+9\st+2\st^2)}{(2+\st)(1+2\st)}z^3\right]\,.
\label{eqn:distz_1d_final}
\end{align}
When $\st=0$, 
the distribution $P(z)$, Eq.~(\ref{eqn:distz_1d_final}), 
is identical to $P(A)$, Eq.~(\ref{eqn:distA_1d_final}), as expected.
In the white-noise limit, the distribution $P(z)$ becomes ($\tilde z=\ku\st z$):
\begin{equation}
P(\tilde z)
=\frac{1}{\sqrt{2 \pi\epsilon^2 }}\left[1-\frac{\tilde z^3}{3\epsilon^2}\right]e^{-\tilde z^2/(2\epsilon^2)}\,,
\end{equation}
where $\epsilon^2=3\ku^2\st$.
This expression is, up to a normalisation factor, identical to a series expansion  of (\ref{eqn:distribution_z_WN}).

Eq.~(\ref{eqn:distz_1d_final}) is compared to results of numerical simulations in Fig. \ref{fig:dists}.
The comparison shows that Eq.~(\ref{eqn:distz_1d_final}) accurately describes the distribution of $z$
at finite but small Kubo numbers provided $z$ is not too large. The expansion employed here
is an expansion in powers of $\ku$ and thus assumes small fluctuations of the random variables ($z$, $A$,$\ldots$).
The large-$z$ behaviour, in particular, cannot be described by this method. This implies that the algebraic
tails of $P(z)$ (related to the formation of caustics) are not captured. The rate of caustic formation
at finite Kubo numbers is computed in the next section, using a WKB approximation.
The results of the present section indicate how Eq.~(\ref{eq:J}) is expected to be
modified at finite Kubo numbers. In the white-noise limit, where Eq.~(\ref{eq:J})
applies, the rate of caustic formation takes an 
activated form [see Eq.~(\ref{eq:J})] when $\epsilon^2 = 3\ku^2 \st$ is small. This is a consequence of the fact that the rate
of caustic formation is determined by the rate of escape of $\tilde z$ from
$\tilde z=0$, as shown by \citet{Meh04}. The typical size of the fluctuations of $\tilde z$ in the vicinity of $\tilde z=0$
is of order $\epsilon^2$, and Eq.~(\ref{eq:J}) assumes the well-known
Arrhenius form obtained from Kramers' theory. At finite Kubo numbers,
the results of this section show that the size of the fluctuations of $\tilde z$
around $\tilde z=0$ are of the order of $\ku^2 \st^2/({1+\st})$.
One therefore expects that the activated rate of caustic formation is given by
(taking $J$ to be positive)
\begin{equation}
\frac{J}{\gamma} \sim  {\rm e}^{-C/(\ku^2 \st^2)}
\label{eq:smallst}
\end{equation}
for small values of $\st$. In the next section we show that this expectation is borne out
for very small values of $\st$ and compute the constant $C$ (it turns out to be $1/96$).
We also compute how the small-$\st$ behaviour (\ref{eq:smallst}) rapidly crosses over to the large-$\st$ behaviour
(\ref{eq:J}).

Finally, we add that in  a similar fashion, the joint distribution of $z$ and $A$ may be obtained:
\begin{align}
P(z,A) &= \frac{(1+\st)}{6\pi\sqrt{\st}} e^{-(A^2-2Az+(1+\st)z^2)(1+\st)/(6\st)}
\bigg[1 +
\frac{\ku}{9(2+\st)(1+2\st)}\Big\{
-18\,\st A+9(-2+3\st+4\st^2)z
\label{eq:PzA}\\
&
-(1+\st)^2(14+13\st+2\st^2)z^3
+ 2A^3(1+\st)
-6(1+\st)(2+\st)A^2z + 3(1+\st)(8+9\st+2\st^2)Az^2
\Big\}
\bigg]
\,.\nonumber
\end{align}
To conclude this section we note that to first order in Kubo number,
and for a Gaussian velocity field $u$ with exponentially decaying
time correlations (as described in section \ref{sec:model}),
the joint distribution of $z$ and $A$ can be obtained from  a Langevin model:
 \begin{eqnarray}
\label{eq:AFP}
 \dot z&=&(A-z)/\st - \ku z^2\,,\quad \dot A=-(A-\overline{A})+F\,.
 \end{eqnarray}
Here $F$ is white noise with zero mean and correlation function $\langle F(t)F(t')\rangle=6\delta(t-t')$.
The Ornstein-Uhlenbeck process for $A$
generates the exponentially decaying time correlations.
Note that $\overline{A}=-3\ku/(1+\st)$ is
added to the fluid velocity gradients in order
to keep the dominant effect of preferential concentration.
We note that the joint distribution of $z$ and $A$ (\ref{eq:PzA}) satisfies 
the Fokker-Planck equation corresponding to Eq.~(\ref{eq:AFP}) to first order in $\ku$.

Eq. (\ref{eq:AFP}) represents the one-dimensional problem of describing the dynamics of
a variable ($z$ in this case) subject to coloured noise in terms of a two-dimensional
white-noise problem. This approach is described in detail by \citet{Wil10} who solved the corresponding
Fokker-Planck equation in perturbation theory.

\section{Rate of caustic formation at finite Kubo numbers}
\label{sec:caustic}
In this section we compute the rate of caustic formation
at finite Kubo numbers. The formation of caustics
is closely associated with the power-law tails
of the distribution $P(z)$. Below we employ
a WKB approximation to determine the rate
of caustic formation. For this calculation it
is convenient to define dimensionless variables
that are slightly different from those used
in section \ref{sec:PzA}: $t = \tilde t/\gamma$,
$x = \tilde x\eta$, $v = \tilde v \eta\gamma$, and $u = \tilde u \eta\gamma$.
These dimensionless variables were employed in \cite{Meh04}
and subsequent work on the white-noise limit of the problem.
In the remainder of this section we drop the tildes
for ease of notation.

Our treatment starts from (\ref{eq:AFP}) which in
the new dimensionless variables takes the form:
\begin{eqnarray}
\label{eq:Langevin}
\dot z&=&A - z - z^2\,,\quad \dot A=-\st (A-\overline{A})+F
\end{eqnarray}
where $F$ is white noise with correlation function 
$\langle F(t)F(t')\rangle=6 \ku^2\st^3\delta(t-t')$.
A comparison between Eqs.~(\ref{eq:AFP}) and (\ref{eq:Langevin}) shows why it is appropriate to
use different dimensionless variables: in this section we expand the dynamics in terms of small $\ku$ 
in small noise levels, whereas in Sec.~\ref{sec:PzA} we expand the dynamics in terms of small $\ku$ 
in the universal contribution due to caustics (the $z^2$-term in (\ref{eq:AFP})).

In the following we omit the non-ergodic correction $\overline{A}$
because we found that its contribution to the rate of caustic formation at the lowest order in $\ku$
is negligible.  In the absence of noise ($F=0$), the dynamics (\ref{eq:Langevin})
has two fixed points. Their coordinates are (setting $\overline{A}=0$):
$(z_1^\ast,A_1^\ast) = (0,0)$ and $(z_2^\ast,A_2^\ast) = (-1,0)$.
The fixed point $(0,0)$ is stable. In the presence of noise
it becomes unstable. But for small noise amplitudes, the variables
$z$ and $A$ fluctuate predominantly in the vicinity of $(0,0)$.
This corresponds to the situation described in Section
\ref{sec:PzA}. However, in the presence of noise the variable $z$
may escape to $-\infty$ via the unstable fixed point $(-1,0)$.
As mentioned in Section \ref{sec:model} this corresponds
to the formation of a caustic. The rate of escape
from the fixed point $(0,0)$ gives the rate of caustic formation.
We note that the Langevin equation (\ref{eq:Langevin})
is of the form considered by \citet{Kane}. Within a
WKB approximation Bray and Kane computed the Kramers escape
rate from a potential well in the presence of coloured noise (with finite correlation
time). The WKB method we outline in the following is
equivalent to that employed in \cite{Kane}.

\subsection{WKB approximation}
\begin{figure}
\psfrag{y}{$A$}
\psfrag{x}{$z$}
\includegraphics[width=6cm]{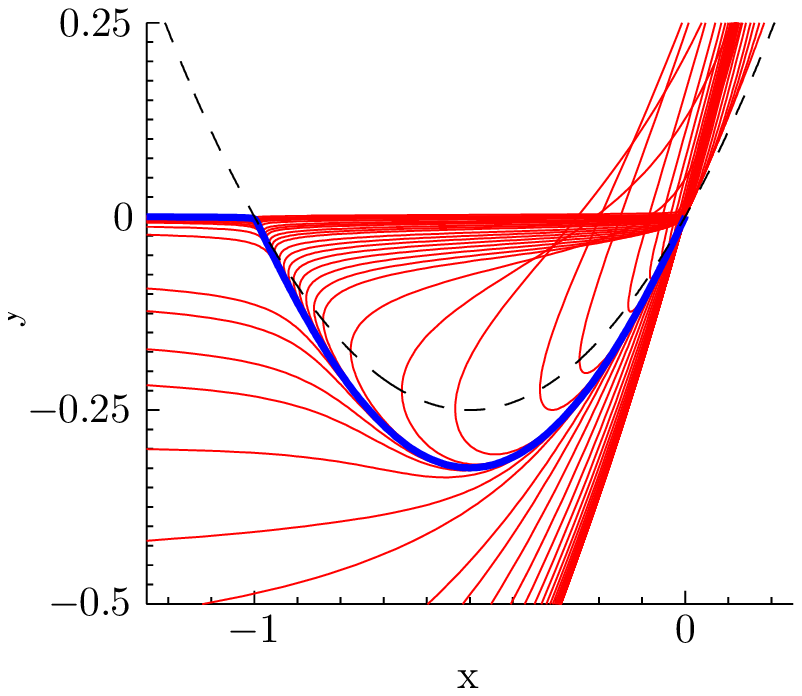}\hspace*{1cm}
\includegraphics[width=6cm]{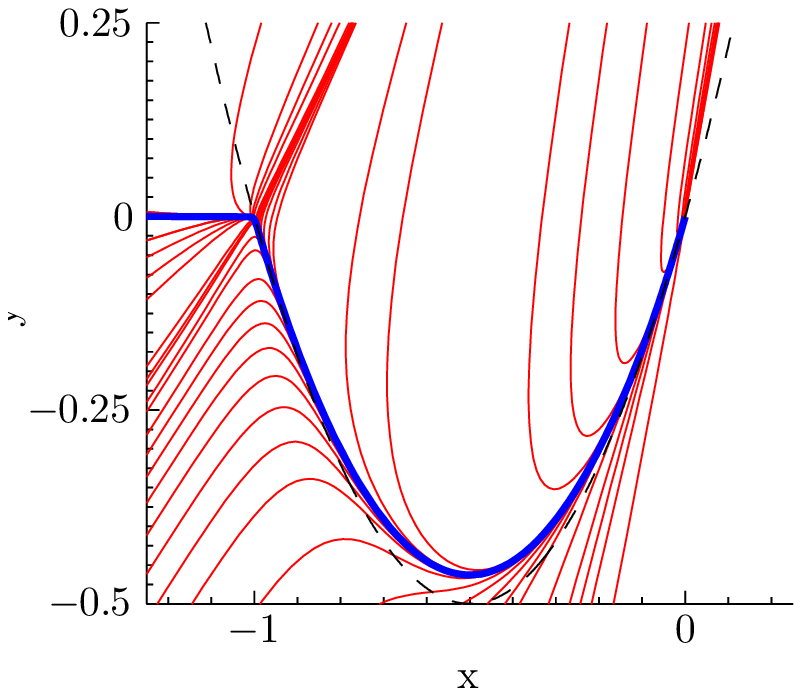}
\caption{\label{fig:trajek} Left: shows Hamiltonian
trajectories in the $z$-$A$-plane for $\st=0.2$, starting
at the fixed point $1$. The optimal trajectory is shown in blue.
The dashed line shows the curve $A=z(z+1)$.
Right: same, but for $\st=2$. Here the dashed line
shows $A=2z(z+1)$.}
\end{figure}
The Fokker-Planck equation corresponding to (\ref{eq:Langevin}) is:
\begin{equation}
\frac{\partial P}{\partial t}=\frac{\partial}{\partial z}
\Big[(z + z^2 - A) P\Big] + \st\frac{\partial}{\partial A} [AP]+
3 \ku^2 \st^3 \frac{\partial^2P}{\partial A^2}\,.
\label{eq:FP_zA}
\end{equation}
In the steady state we have $\partial P/\partial t= 0$ and
we seek a solution of the steady-state Fokker-Planck equation
of the form
\begin{equation}
\label{eq:ansatz}
P(z,A)=\exp[-S(z,A)/\ku^2 + \mbox{higher orders in $\ku$}]
\end{equation}
(for a comprehensive description of the WKB method discussed
in the following paragraphs
see \citet{Dyk94} and references cited therein).
The function $S(z,A)$ is referred to as the \lq action'.
It is defined such that $S=0$ at the fixed point $(0,0)$.
We expect $S$ to be a quadratic function of $z$ and $A$ in the vicinity of this point, corresponding to the
case considered in Section \ref{sec:PzA}. Here by contrast  we
are interested in the tails of the distribution, corresponding
to large deviations of $(z,A)$ from $(0,0)$.
Inserting the ansatz (\ref{eq:ansatz}) into the
Fokker-Planck equation and
expanding in powers of $\ku$,
one obtains a first-order partial differential equation for $S$:
\begin{equation}
\big(A-z(1+z)\big) \frac{\partial S}{\partial z}-\st A \frac{\partial S}{\partial A}+3\st^3 \Big(\frac{\partial S}{\partial A}\Big)^2=0\,.
\label{eq:WKB0}
\end{equation}
This equation has the form of a Hamilton-Jacobi equation
$H(\ve q, \ve p) = 0$, with coordinates $\ve q = (z,A)^{\sf T}$
and \lq momenta' $\ve p = (p_z,p_A)^{\sf T}$ with
$p_z = \partial S/\partial z$ and $p_A = \partial S/\partial A$.
We write the \lq Hamiltonian' in the standard form (see \cite{Wolff}
and references cited therein):
\begin{equation}
\label{eq:Hst}
H = \ve p^{\sf T}\ve v + {1\over 2} \ve p^{\sf T} {\bf D} \ve p\,,
\end{equation}
with
\begin{equation}
\ve v = \left(\begin{array}{l} A-z(1+z)\\ -\st \,A
\end{array}\right)\,,\quad\mbox{and}\quad {\bf D} = \left(\begin{array}{ll}
0 &0\\0 &6\,{\rm St}^3\end{array}\right)\,.
\end{equation}
The solution of Eq.~(\ref{eq:WKB0}) is found by solving
the Hamiltonian dynamics corresponding to (\ref{eq:Hst}):
\begin{equation}
\label{eq:Heq}
\dot{\ve q} = \partial H /\partial \ve p\,,\quad
\dot{\ve p} = -\partial H/\partial \ve q\,.
\end{equation}
These equations exhibit three fixed points $(\ve q^\ast,\ve p^\ast)$
given in Table \ref{tab:FP}. The first one $(\ve q_1^\ast,\ve p_1^\ast) = (0,0,0,0)$
corresponds to the stable fixed point of the noise-less dynamics
discussed above. The second fixed point $(\ve q_1^\ast,\ve p_1^\ast) = (-1,0,0,0)$
corresponds to the saddle of the noise-less dynamics.
Consider solutions of (\ref{eq:Heq}) satisfying
\begin{equation}
\label{eq:bc}
\left . \begin{array}{c}
\ve q(t) \rightarrow  \ve q_1^\ast \\
\ve p(t) \rightarrow \ve p_1^\ast
\end{array}
\right \} \quad\mbox{as $t\rightarrow-\infty$}\,,
\quad\ve q(t) \rightarrow \ve q\quad\mbox{as $t\rightarrow \infty$}\,,
\quad \mbox{and $H(\ve q,\ve p)=0$}\,.
\end{equation}
To every such solution corresponds an action $S(\ve q)$:
\begin{equation}
S(\ve q) = \int_{-\infty}^\infty {\rm d}t \ve p \,\dot{\ve q}\,.
\end{equation}
\citet{Frei98} gave a variational principle for the most likely path
of escape satisfying the boundary conditions (\ref{eq:bc}). In the limit
of small values of $\ku$, the probability distribution
(\ref{eq:ansatz}) is dominated by the escape path with extremal action.
\begin{table}[t]
\begin{tabular}{lcccc}
\hline\hline
&$z^\ast$ & $A^\ast$ & $p_z^\ast$ & $p_A^\ast$\cr\hline
$1$&$0$ & $0$ & $0$ & $0$\cr
$2$&$-1$ & $0$ & $0$ & $0$\cr
$3$&$-\frac{1}{2}$ & $-\frac{1}{4}$ & $-\frac{1}{24\st}$ & $-\frac{1}{24\st^2}$\\[1mm]\hline\hline
\end{tabular}
\caption{\label{tab:FP}
Fixed points of the Hamiltonian dynamics (\ref{eq:Heq}).}
\end{table}

Fig. \ref{fig:trajek} shows numerical solutions of Eqs. (\ref{eq:Heq})
with the boundary conditions (\ref{eq:bc}) for two
different values of the Stokes number. Shown are trajectories
in the $z$-$A$-plane. These trajectories must leave
the fixed point $(\ve q_1^\ast,\ve p_1^\ast)$ along one of its unstable directions.
These are found by linearising Hamilton's equations in the vicinity
of $(\ve q_1^\ast,\ve p_1^\ast)$:
\begin{equation}
{\bf J}_1 =
\left(\begin{array}{cc}
{\bf A}_1 & {\bf D}\\
{\bf 0} & -{\bf A}_1^{\sf T}\end{array}\right)\,.
\end{equation}
Here ${\bf A}_1$ is the linearisation of the
noise-less dynamics $\dot{\ve q} = \ve v$ with elements
$A_{ij} = \partial v_i/\partial q_j$ evaluated at $(\ve q_1^\ast,\ve p_1^\ast)$:
\begin{equation}
{\bf A}_1 =
\left(
\begin{array}{cc}
-1 & 1\\
0 & -\st
\end{array}\right)\,.
\end{equation}
The linearised Hamiltonian dynamics in the vicinity of
$(\ve q_1^\ast,\ve p_1^\ast)$ satisfies
\begin{equation}
\label{eq:dp}
\delta \ve p = {\bf C}^{-1} \delta \ve q \quad\mbox{with}\quad
{\bf C}^{-1} =
\frac{1+\st}{\st^3}
\left(
\begin{array}{cc}
 1+\st&-1 \\
 -1&  1
\end{array}\right)\,,
\end{equation}
and gives rise to the action
\begin{eqnarray}
\label{eq:action0}
S(\ve q)&\approx&\frac{1}{2} \ve q^{\sf T} {\bf C}^{-1} \ve q
= \frac{1 + \st}{6\st^3} (A^2 - 2 A z + (1 + \st) z^2)\,.
\end{eqnarray}
We see that this result is consistent with Eq.~(\ref{eq:PzA}),
and note that the additional factor $\ku^2\st^2$ is a consequence
of the fact that different dimensionless variables
are used in Sections \ref{sec:PzA} and \ref{sec:caustic}.

Eq. (\ref{eq:action0}) [or alternatively Eq. (\ref{eq:dp})]
constrains the initial conditions of the Hamiltonian dynamics,
leaving only one parameter to be varied, the angle in the $z$-$A$-plane infinitesimally close to the first fixed point.
Fig. \ref{fig:trajek}
shows the corresponding families of trajectories.
The \lq optimal escape path' with extremal action
is shown as a blue solid line. Its form (and the corresponding
value of the action) depends upon the Stokes number.
For large Stokes numbers, the optimal path
approaches the curve
\begin{equation}
\label{eq:path_large_st}
A = 2z(1+z)
\end{equation}
in the $z$-$A$-plane. In the limit of $\st \rightarrow \infty$,
the $z$ dynamics thus approaches $\dot z = z + z^2$. This is the equation
of motion determining the optimal escape path in
the white-noise limit \cite{Meh04}. In the limit $\st \rightarrow 0$,
by contrast, the optimal path approaches the curve given by
\begin{equation}
A = z(1+z)\,.
\label{eqn:cond_small_St}
\end{equation}
This condition corresponds to $\dot z = 0$ (in terms of the dimensionless units adopted in Sec.~\ref{sec:PzA},
the condition (\ref{eqn:cond_small_St}) corresponds to advection, $z=A$).
\begin{figure}[t]
\psfrag{y}{\large -$\ku^2 \log J/\gamma$}
\psfrag{x}{\large $\st$}
\includegraphics[width=10cm]{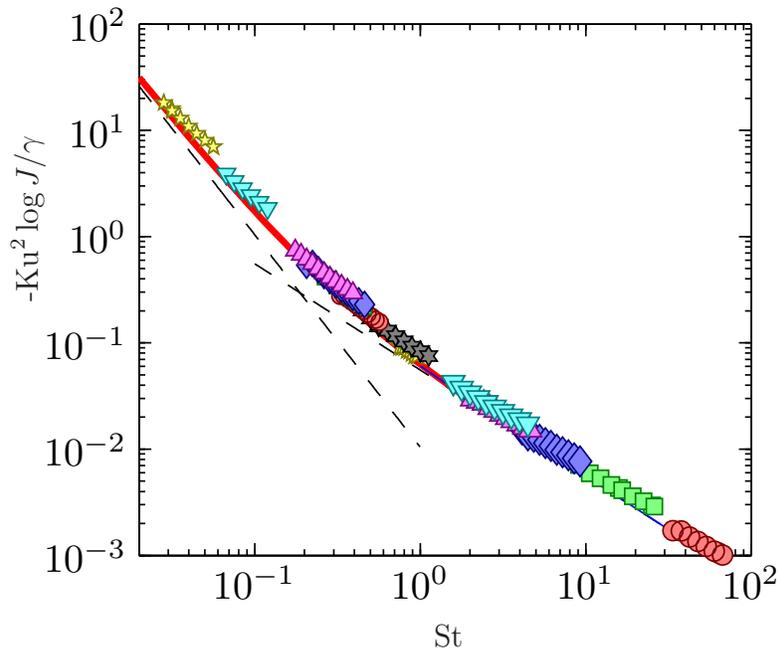}
\caption{\label{fig:action}
Shows the action $S$ in (\ref{eq:result}) as a function of $\st$ (solid red line).
Also shown are the limiting behaviours (\ref{eq:S_large_st}) for large values of $\st$ and (\ref{eq:S_small_st}) for small values of $\st$
(dashed lines). Finally, results of numerical simulations of the model described in Section \ref{sec:model} are shown (symbols).
In order to allow for a quantitative comparison with the WKB theory, only parameter combinations
that give rise to $J/\gamma < 10^{-3}$ are shown.
Parameters:
$\ku=0.01$ (red,$\circ$), $0.02$ (green,$\Box$), $0.03$ (blue,$\Diamond$), $0.04$ (magenta,$\vartriangle$), $0.05$ (cyan,$\triangledown$), $0.07$ (yellow,$\star$), $0.1$ (black,$\ast$), $0.12$ (red,small $\circ$), $0.15$ (green,small $\Box$), $0.17$ (blue,small $\Diamond$), $0.2$ (magenta,small $\vartriangle$), 
$0.5$ (cyan,small $\triangledown$), and $1$  (yellow,small $\star$).
 }
\end{figure}
Given the optimal value $S$ of the action,
the rate of caustic formation is given by
\begin{equation}
\label{eq:result}
\frac{J}{\gamma} \sim {\rm e}^{-S/\ku^2}\,.
\end{equation}
Our results for the action are summarised in Fig.~\ref{fig:action}.
This figure shows results for $-\ku^2 \log J/\gamma$ (symbols)
obtained from numerical simulations of the model described
in Section \ref{sec:model}. When $J/\gamma$ is small,
this expression is approximately given by the action.
In the figure, results for different Stokes and
Kubo numbers are plotted, subject to the condition $J/\gamma < 10^{-3}$.
Also shown is the action corresponding to the optimal
escape path found by numerically integrating (\ref{eq:Heq}), solid red line.
We observe good agreement for Kubo numbers up to $\ku \sim 1$,
despite the fact that the WKB-approximation is a small-$\ku$ approximation.
The asymptotic behaviours for small and large values of $\st$
(shown as dashed lines in Fig.~\ref{fig:action}) are discussed
in the following Subsection.

\subsection{Perturbation theory}
We have found an analytical expression for the action $S(\ve q)$ which turns out to give the correct escape action for $\st$ larger than $\st\sim1$.  This expression is derived
by expanding the action around $\ve q=\ve 0$:
\begin{equation}
S(\ve q)=\sum_{i=0}^\infty S^{(i)}(\ve q),\mbox{ where }S^{(i)}(\ve q)=\sum_{j=0}^i a^{(i)}_{j}z^jA^{i-j}\,,
\label{eqn:action_ansatz}
\end{equation}
where $a^{(i)}_{j}$ are the expansion coefficients to order $i$.
Provided the series (\ref{eqn:action_ansatz}) converges,
we may determine the action at the second fixed point (and thus the action of the trajectory escaping to $z=-\infty$)
by evaluating (\ref{eqn:action_ansatz}) at $z=-1$ and $A=0$, i.e. $S=\sum_{i=0}^\infty(-1)^ia^{(i)}_{i}$.
From Eq. (\ref{eq:action0}) we know the action close to $z=A=0$ which immediately gives $S^{(i)}(\ve q)$ for $i\le 2$.
We determine the higher orders $S^{(i)}(\ve q)$ by inserting
(\ref{eqn:action_ansatz}) into (\ref{eq:WKB0}), requiring that all terms of order $i$ vanish.
In this way we obtain the recursion
\begin{align}
a^{(i)}_{j+1}
=
\frac{1}{j+1}
\left[
a^{(i)}_{j}(j+\st(i-j))
+a^{(i-1)}_{j-1}(j-1)
-3\st^3\sum_{m=2}^{i}\sum_{n=0}^m a^{(i-m+2)}_{j-n}a^{(m)}_{n}(i-j-m+n+2)(m-n)
\right]
\end{align}
for $j=0,\dots,i-1$. Together with
\begin{align}
a^{(i)}_{i}i
+a^{(i-1)}_{i-1}(i-1)
-3\st^3\sum_{m=2}^{i}\sum_{n=0}^{m-1} a^{(i-m+2)}_{i-n}a^{(m)}_{n}(-m+n+2)(m-n)=0
\end{align}
this recursion
determines all coefficients to order $i$.
To third order we find
\begin{align}
&a^{(3)}_0=-\frac{2(1+\st)}{9\st^3(2+\st)(1+2\st)}\,,
&&a^{(3)}_1=\frac{2(1+\st)}{3\st^3(1+2\st)}\,,\nn\\
&a^{(3)}_2=-\frac{8+17\st+11\st^2+2\st^3}{3\st^3(2+\st)(1+2\st)}\,,
&&a^{(3)}_3=\frac{(1+\st)^2(14+13\st+2\st^2)}{9\st^3(2+\st)(1+2\st)}\,.
\label{eqn:perturb_coeffs_3}
\end{align}
From these coefficients $S^{(3)}$ may be calculated.

A number of comments are due at this point.  First, we note that expanding the action obtained in this way in powers of $\st^{-1}$ corresponds to the small-$\tau$ expansion of the escape action obtained by \citet{Kane}.

Second, expanding to lowest order in $z^jA^{3-j}$ we find
\begin{equation}
P(z,A) \sim {\rm e}^{-(S^{(2)}+S^{(3)})/\ku^2}
\sim {\rm e}^{-S^{(2)}/\ku^2}(1-S^{(3)}/\ku^2)\,.
\end{equation}
The last expression corresponds to the small-$\ku$ limit of (\ref{eq:PzA}).
Factors of $\ku\st$ arises because different
dimensionless units are used in Section \ref{sec:PzA} and here.

Third, expanding the action to $S^{(2)}+S^{(3)}$ at $z=-1$ and $A=0$
to lowest order in  $\st^{-1}$ results in $S=1/(18\,\st)$. This is again the white-noise result (\ref{eq:J}).

Fourth, expanding the higher-order coefficients $a^{(j)}_{j}$ in $\st^{-1}$ gives
\begin{equation}
S = \frac{1}{18\, \st} + \frac{1}{90\, \st^3} - \frac{1}{105\, \st^5} + \frac{4}{189\,\st^7}+\ldots\,.
\label{eq:S_large_st}
\end{equation}
This result can also be obtained from  the lowest order of the small-$\tau$ expansion by \citet{Kane}
(Eq.~(12) in their paper).  The lowest order of this expression
is the white-noise result, shown as a dashed line in Fig.~\ref{fig:action}.

The small-$\st$ asymptotics of the action can be derived
following a procedure outlined in \cite{Kane}.
The starting point is Eq.~(\ref{eqn:cond_small_St}).
This condition can be interpreted as a fixed-point condition
for the $z$-dynamics which rapidly adjusts to the slowly-moving
variable $A$.  This fixed point becomes unstable for $z<-1$ (and for $z>0$).
In the former case, the $z$-coordinate escapes
from $z=-1$ to $z=-\infty$ with $A=0$.
Solving Eq.~(\ref{eqn:cond_small_St}) for $z$ in the range $-1<z<0$ gives $z^{(\pm)}=-1/2\pm\sqrt{1+4A}/2$.
The solution $z^{(-)}$ is appropriate in the interval $-1<z<-1/2$,
while the solution $z^{(+)}$ pertains to  $-1/2<z<0$.
There are two corresponding
solutions of Hamilton's equations subject to the
condition (\ref{eqn:cond_small_St})
\begin{align}
p^{(+)}_z&=0\,, & p^{(+)}_A&=A^{(+)}_0e^{\st t}/(3\st^2)\,, & A^{(+)}&=A^{(+)}_0e^{\st t}\,,\\
p^{(-)}_z&=0\,, & p^{(-)}_A&=0\,, & A^{(-)}&=A^{(-)}_0e^{-\st t}\,.
\end{align}
The first solution (denoted $(+)$)
describes how  $A$ decreases
from a small negative value $A^{(+)}_0$ at $z=0$ to $A=-1/4$ at $z=-1/2$.
Then the second solution (denoted by $(-)$)
takes over, describing how  $A$ increases
to $A=0$ at $z=-1$ as $t\to\infty$.
The action vanishes for the $(-)$-branch, and the action for
the $(+)$-branch gives:
\begin{equation}
\label{eq:S_small_st}
S = \frac{1}{96\, \st^2}\,.
\end{equation}
This result is equivalent to the lowest order
in a large-$\tau$ expansion given in \cite{Kane}.
Eq.~(\ref{eq:S_small_st}) is shown as a dashed line
in Fig.~\ref{fig:action}. Our numerical results
(solid red line) converge to this asymptote,
albeit very slowly.

\section{Conclusions}
\label{sec:conc}
In this paper we have computed the distribution
of particle-velocity gradients $z = \partial v/\partial x$
and the rate of caustic formation at finite Kubo numbers,
for a one-dimensional model of a turbulent aerosol.
The body of the distribution was computed by means
of a perturbation expansion in Kubo number, and the
rate of caustic formation within a WKB approximation.
We have described the differences between these two approaches,
and how they are consistent. The results complement and extend
results obtained earlier in the white-noise limit
$\ku\rightarrow 0$ and $\st \rightarrow \infty$, such that
$\epsilon^2 = 3\ku^2 \st$ remains constant.
These earlier results have yielded important insight into
the mechanisms governing inertial particle dynamics,
but it is of interest to ask how these results are
modified at finite Kubo numbers - since turbulent
flows have Kubo numbers of order unity.

A finite Kubo number has two effects: the fluid velocities
have a finite correlation time, and the suspended particles may explore
configuration space preferentially. Both affect
the distribution of particle-velocity gradients and the rate of caustic formation.
To lowest order in $\ku$, modifications are due to the finite correlation time of the flow.
The effect due to preferential sampling occurs at higher order in $\ku$.

We find that the rate of caustic formation exhibits an activated
$\st$-dependence, as in the white-noise limit.
For $\st > 1$, our result for the rate of caustic formation agrees
with earlier results obtained in the
white-noise limit:
\begin{equation}
\label{eq:result1}
\frac{J}{\gamma} 
\sim {\rm e}^{-1/(6 \epsilon^2)}\quad\mbox{with $\epsilon^2 = 3\ku^2 \st$}\,.
\end{equation}
For very small Stokes numbers ($\st < 0.04$) we find,
by contrast, that
\begin{equation}
\label{eq:result2}
\frac{J}{\gamma} 
\sim {\rm e}^{-1/(96\ku^2 \st^2)}\,.
\end{equation}
This scaling is consistent with a parameterisation of the collision
rate for particles in turbulent aerosols proposed in \cite{Fal02}.
At intermediate Stokes numbers ($0.04 < \st < 1$ for the model
considered here) the $\st$-dependence of the action 
is more complicated, but the WKB approximation yields a quantitative
description. The action for the caustic-rate formation
is shown in Fig. \ref{fig:action}. The crossover between 
the two asymptotic expressions (\ref{eq:result1}) and (\ref{eq:result2}) 
occurs at $\st = 3/16$ for the model considered here. 

While we expect essentially similar results in two and
three spatial dimensions it is nevertheless important to check this
in detail. A difference between one and higher dimensions is of course
that one-dimensional flows are always compressible.
Using the approach described in Section \ref{sec:model} it
will be possible to compute the probability that particles
are found in vortical regions in two- and three-dimensional
incompressible flows at finite Kubo numbers.
This is of interest since this probability was measured
in direct numerical simulations of inertial particles
suspended in turbulent flows \cite{Bec07}.

In one spatial dimension, a number of open questions still remain.
First, the WKB approximation was only performed to lowest
order in $\ku$, the prefactor of the exponential in the distribution of $z$ must also be computed. It is expected
that it will give rise to power-law tails $\sim |z|^{-2}$
of the distribution of $z$. Second, it remains to be seen to which
extent preferential sampling at finite Kubo numbers modifies
this prefactor.
\\[0.2cm]
{\em Acknowledgements.} We gratefully acknowledge
financial support by Vetenskapsr\aa{}det and
by the G\"oran Gustafsson Foundation for Research in
Natural Sciences and Medicine.

\bibliography{biblio}

\end{document}